\documentclass[twocolumn, apj]{emulateapj}
\usepackage{natbib, amsmath}

\newcommand{\dif}{\mathrm{d}}

\def\Mo               {\hbox{$M_{\odot}$}}

\def\om{\omega}
\def\omt{\tilde \omega}
\def\omc{\omega_{\rm cond}}
\def\bdotr{$\langle|\boldsymbol{\hat{b}}\cdot\boldsymbol{\hat{r}}|\rangle$}

\shorttitle{Magnethothermal Instability in Clusters}
\shortauthors{Parrish, Stone, \& Lemaster}

\begin{document}
\title{The Magnetothermal Instability in the Intracluster Medium}
\author{Ian J. Parrish\altaffilmark{1,2,3}, James M. Stone\altaffilmark{3}, and Nicole Lemaster\altaffilmark{3}}
\altaffiltext{1}{Astronomy Department \& Theoretical Astrophysics Center, 601 Campbell Hall, The University of California, Berkeley, CA 94720; iparrish@astro.berkeley.edu}
\altaffiltext{2}{Chandra Fellow}
\altaffiltext{3}{Department of Astrophysical Sciences, Peyton Hall, Ivy Lane, Princeton University, NJ 08544}

\begin{abstract}
The electron mean free path in the intracluster medium (ICM) of galaxy clusters is much larger than the gyroradius; thus, heat is transported anisotropically along magnetic field lines.  We show that the intracluster medium is unstable to the magnetothermal instability (MTI) using MHD simulations with anisotropic thermal conduction.  As a result of the MTI, we find that the temperature profile of the ICM can be substantially modified on timescales of several billion years while the magnetic field is amplified by dynamo action to more than fifty times the original energy.  We also show that the instability drives field lines to become preferentially radial leading to conduction that is a highly efficient fraction of the Spitzer conductivity.  As such, we present the first self-consistent calculation of the effective thermal conductivity in the ICM.
\end{abstract}
\section{Introduction}\label{sec:intro}
Clusters of galaxies are the most massive gravitationally-bound objects in the universe, yet a mere 3\% of their mass is in the form of stars.  The majority of their mass (84\%) is in dark matter with the remaining 13\% consisting of a hot, low density, magnetized plasma called the intracluster medium (ICM). Due to their enormous masses, surveys of clusters provide key tests of cosmology.  With the Chandra X-ray Observatory, we are capable of measuring the spatially resolved X-ray luminosity ($10^{43}$--$10^{46}$ erg/s) emitted by the ICM to determine its density as a function of radius.  In order to fully understand the fundamental astrophysics of the largest bound objects in the universe, we must develop an accurate quantitative picture of the heating and cooling mechanisms as well as the dynamical processes that govern the ICM.   Thus, we are motivated by the use of clusters as a cosmological tool to study the physics of the ICM much more rigorously.

The intracluster medium is heated by gravitational infall such that typical temperatures are 1--15 keV and typical densities are in the range of $10^{-3}$--$10^{-2} \;\mbox{cm}^{-3}$ \citep{pf06}.  Since the ICM is fully ionized and radiative forces are negligible, we can treat radiation with optically thin cooling and neglect electrical resistivity.  Estimates for the magnetic field strength in clusters range from roughly 1--10~$\mu$G at the center and 0.1--1.0~$\mu$G at a radius of 1 Mpc, values that correspond to a plasma beta, $\beta = 8\pi P/B^2\approx 200$--$2000$, a dynamically weak magnetic field \citep{ct02}.  In this dilute plasma, however, the mean free path for electron collisions can be twelve orders of magnitude larger than the gyroradius \citep{nm01}.  In this circumstance, the equations of MHD must be supplemented with anisotropic terms that include the near free-streaming motions of particles along magnetic field lines \citep{brag65}.  The parallel thermal conductivity of the electrons is larger than that of the ions by a factor proportional to $(m_i/m_e)^{1/2}$, whereas the parallel viscosity of the ions is larger than that of the electrons by the same factor.  \citet{bal00} has predicted that gravitationally bound plasmas in this regime are susceptible to the magnetothermal instability (MTI).  This convective instability is driven by anisotropic thermal conduction along field lines with major dynamical consequences \citep{ps05,ps07a,ps07b,cd06}.  Thus, we are no longer justified in considering the weak magnetic field in the ICM as a purely passive component in a fluid governed solely by hydrodynamics.

There are two long-standing puzzles about the ICM.  First, the radial temperature profiles show a slight decline in temperature with radius \citep{pratt07} that is smaller than that expected from structure formation calculations, e.g. \citep{lok02}.  Second, the magnetic field has been significantly amplified in the cluster from its primordial value of $< 1$ nG.  The ICM is unstable to the magnetothermal instability on scales of tens of kpc and larger, as calculated from the current observed magnetic field strengths. In three-dimensional simulations, we have shown that the MTI is capable of generating convective motions and a magnetic dynamo while efficiently transporting heat.  In addition, the MTI can make the temperature profile more isothermal as it exhausts its source of free energy.  These similarities provide a strong motivation for studying the ICM with this type of simulation.  

Hydrostatic equilibrium dictates that
\begin{equation}
M(r) = -\frac{k_B r^2}{G\mu m_p}\left(T_e(r)\frac{\dif n_e(r)}{\dif r} + n_e(r)\frac{\dif T_e(r)}{\dif r}\right),
\label{eqn:clust:m-cluster}
\end{equation}
where $\mu$ is the mean molecular weight \citep{laroque06}.
When an X-ray observation is made, the density profile (first term of eqn. [\ref{eqn:clust:m-cluster}]) can be constrained fairly well due to the property that $L_x \propto n_e^2$ for free-free emission, the dominant cooling mechanism. Often, the gas density profile is fit with an isothermal-$\beta$ model \citep{cff76}. On the other hand, the temperature profile (second term of eqn. [\ref{eqn:clust:m-cluster}]) can only be constrained by spectra, which are increasingly difficult to measure at larger redshifts.    For high redshift clusters, the virial masses are often constrained simply by assuming isothermality, in contradiction to local observations. Why do we care about determing mass with such accuracy?  A flux-limited X-ray survey can be used to constrain $\Omega_0$, the matter parameter, e.g. \citet{borg99}.  Locally, however, there is a degeneracy between the parameters $\Omega_0$, the matter fraction of the universe, and $\sigma_8$, the variance of the power spectrum on $8 h^{-1}$ Mpc scales.  Thus, one must examine the evolution of the cluster mass function with redshift to break this degeneracy.  

In \S\ref{sec:MTIphys}, we review the physics of the MTI then procede to construct a model of the ICM for numerical simulation in \S\ref{sec:model}.  In \S\ref{sec:method} we outline our numerical methods.  We then show in \S\ref{sec:sim} that the MTI is capable of rearranging the temperature profiles of clusters on cosmologically important timescales, \textit{i.e.} 0.5--10 billion years.  We also show that a magnetic dynamo operates as a result of the turbulent convection driven by the MTI.  A comparison to runs with purely isotropic conduction at a specified fraction of the Spitzer conductivity show that much less turbulence and no dynamo result.  Finally, we point out that the magnetic field is driven to a highly radial geometry and speculate on possible observational implications before concluding.

\section{Physics of the MTI}\label{sec:MTIphys}
The physics of the magnetothermal instability are clearly outlined in \citet{bal00} and \citet{quat08}.  We will only briefly review them here.  The magnetothermal instability is a convective instability in which the source of free energy is the temperature gradient.  In the more familiar Schwarschild convection, the entropy gradient is the source of free energy.  For an arbitrarily oriented magnetic field, the dispersion relation of the MTI can be given by \citet{quat08}:\begin{eqnarray} \label{eqn:DR} &0& =  \om \omt^2
 + i \omc \omt^2 - N^2 \omega {k_\perp^2 \over k^2} \nonumber \\ &-& i
 \omc g \left({d\ln T \over dz}\right) \left[(1 - 2b_z^2){k_\perp^2
 \over k^2} + {2b_xb_zk_xk_z \over k^2}\right], 
\end{eqnarray}
where $N$ is the Brunt-V\"ais\"al\"a frequency,
\begin{equation}
N^{2} = -\frac{1}{\gamma \rho}\frac{\partial P}{\partial z}\frac{\partial \ln S}{\partial z},
\label{eqn:brunt-vaisala}
\end{equation}
and
\begin{equation}
\omt^2 = \om^2 - \left(\boldsymbol{k}\cdot\boldsymbol{v_{A}}\right)^2,
\label{eqn:omegatilde}
\end{equation}
where $\boldsymbol{v_A} = \boldsymbol{B}/(4\pi\rho)^{1/2}$ is the Alfv\'{e}n speed, and finally
\begin{equation}
\om_{\textrm{cond}} = \frac{2}{5}\chi\left(\boldsymbol{\hat{b}}\cdot\boldsymbol{k}\right)^2
\label{eqn:omegacond}
\end{equation}
is the frequency for conduction to act on a given scale with $\boldsymbol{\hat{b}}$ the unit vector directed along the magnetic field and $\chi$ the thermal diffusivity\footnote{The literature is not consistent as to the use of $\chi$ and $\kappa$.  This work will use $\chi$ to represent a true diffusion coefficient and $\kappa$ to represent a conductivity in units erg cm$^{-1}$ s$^{-1}$ K$^{-1}$.} in units of cm$^2$s$^{-1}$.  This dispersion relation is written without loss of generality for the geometry in which gravity and the initial atmospheric gradients are in the $\boldsymbol{\hat{z}}$-direction and the initial magnetic field lies in the $\boldsymbol{\hat{x}}$--$\boldsymbol{\hat{z}}$ plane. 

We now move to the limit of the MTI in which the temperature decreases with height in the atmosphere and the magnetic field is initially perpendicular to $\boldsymbol{g}$.  In this limit the growth rate of the instability simplifies  to
\begin{equation}
\om^2 \approx g\left(\frac{\dif \ln T}{\dif z}\right)\frac{k_{\perp}^2}{ k^2},
\label{eqn:omegasimple}
\end{equation}
which is growth at the isothermal limit of the Brunt-V\"ais\"al\"a frequency, where $k_{\perp}$ is defined as being perpendicular to $\boldsymbol{g}$. The MTI can be stabilized by strong magnetic fields, when magnetic tension begins to retard the convective motions, or by isotropic conduction that short-circuits the anisotropic driving necessary for this instability.  \citet{ps05} and \citet{ps07b} describe local simulations of the growth and saturation of this instability in two- and three-dimensions.  In short, the MTI saturates through a combination of three processees.  First, the instability drives a magnetic dynamo that amplifies the field, increasing magnetic tension.  Second, if the system has adiabatic boundaries such that the temperature gradient can be decreased significantly, the system approaches isothermal and the driving free energy term vanishes.  Finally, in what perhaps is the most surprising result, the MTI reorients the direction of the magnetic field, preferentially aligning it with the background temperature gradients.  This realignment reduces the growth rate of the instability as the field becomes more radial (aligned with the temperature gradient) and can be seen by examining equation (\ref{eqn:DR}).  We shall see all of these saturation behaviors present in our study of the MTI in galaxy clusters.

\section{Cluster Model}\label{sec:model}
%See notes circa 5/5/07
Before beginning any simulations of a cluster, the first step is to choose an appropriate model for the dark matter and ICM that is inspired by observations and theoretically consistent.  Unfortunately, there is a bit of a ``chicken and egg problem".  What we are observing today is potentially already the saturated state that results from whatever instabilities and processes have been taking place since the formation of the cluster.  Nonetheless, we shall choose models that are generally consistent with the profiles observed today to study the general properties of the ICM, as opposed to modeling any one specific cluster.

Much of the inspiration for our cluster model comes from \citet{hughes89}, in which the mass of the Coma Cluster is estimated from X-ray observations.  Several ingredients go into the model which we shall discuss in turn: a profile for the dark matter, a profile for the gas properties, and a way to tie them together.  Since major mergers are quite infrequent, the latter is quite easy to achieve, simply by assuming hydrostatic equilibrium.  
\subsection{Dark Matter and Gas Profiles}\label{subsec:profiles}
First, the dark matter profile:  at the time of Hughes' paper, there was almost universal agreement that dark matter in clusters was distributed according to the King dark matter profile \citep{king66}.  The profile is parameterized as
\begin{equation}
\rho_{dm}(r) = \rho_{dm 0}\left[1 + \left(\frac{r}{r_{dm}}\right)^2\right]^{-n/2},
\label{eqn:clust:king}
\end{equation}
where $\rho_{dm}$ is the dark matter density binding the cluster, $r_{dm}$ is the scale radius, and $n$ = 3, 4, or 5 in general.  In the last decade simulations and some observational evidence have favored a somewhat cuspy central region given by the so-called NFW models \citep{nfw97}
\begin{equation}
\frac{\rho_{dm}(r)}{\rho_{crit}} = \frac{\delta_c}{(r/r_s)(1+r/r_s)^2},
\label{eqn:clust:nfw}
\end{equation}
where $\delta_c$ is the halo overdensity compared to the critical density at that epoch, and $r_s$ is the NFW scale radius.  This work adopts the King model for the dark matter profile for several reasons.   To begin with, we are not attempting to model the central region in detail, so the central region is less relevant.  In addition, the cuspy dark matter profile makes it more difficult to hold the cluster in quiescent hydrostatic equilibrium.  Finally, for the choice of $n=3$ both profiles have $\rho_{dm}\propto r^{-3}$ on scales larger than the core radius, and thus are roughly equivalent.  This work will use the $n=3$ King model for dark matter throughout.  

Again for the gas profile (the terms ``gas'' and ``ICM'' are used interchangeably in the literature) there are several possible choices.  Hughes' paper makes the common choice of the isothermal beta model \citep{cff76}
\begin{equation}
\rho_g(r) = \rho_{g0}\left[1+\left(\frac{r}{r_c}\right)^2\right]^{-3\beta/2},
\label{eqn:clust:betamodel}
\end{equation}
where $\rho_g$ is the gas density, $r_c$ is the gas scale radius, and the $\beta$ parameter is of order 0.75.  \citet{hughes89} derives cluster models by combining the King model with the isothermal beta model.  There are two primary shortcomings to this method.  First, the resultant temperature profiles are incredibly sensitive to the various free parameters, yielding many models that have unphysical temperatures that go to zero or infinity quite rapidly.  Second, these models are in general not convectively stable everywhere according to the Schwarzschild criterion, a poor initial condition for understanding a convective instability.  It is also worth noting that these models arose during a time when X-ray measurements did not have sufficient resolution to resolve temperature gradients in clusters; thus, they are not necessarily as physically motivated today.  

An appealing alternate model for the gas density is specified through a fit to the entropy profile.  More recent X-ray observations are well fit by such profiles \citep{donahue06}.  The general model is of the form
\begin{equation}
S(r) = S_0\left[1 + \left(\frac{r}{r_c}\right)^{\alpha}\right],
\label{eqn:clust:gas-entropy}
\end{equation}
where $S= p\rho^{-\gamma}$ is the gas entropy, $r_c$ is the entropy scale radius, and $\alpha$ is a parameter.  For $\alpha > 1$, $\dif S/\dif z>0$ and the atmosphere is manifestly convectively stable.  Outside of the core many clusters are well fit observationally by $\alpha = 1.3$, thus, that is the fiducial choice for this work.  In general for the gas profiles there are three thermodynamic quantities: density, pressure, and temperature.  One of these is determined by the equation of state, one of them is specified as an \textit{Ansatz}, and the third must be calculated.  When using the entropy model, the entropy is specified instead of one of these quantities.   
\subsection{Generating Self-Consistent Cluster Models}\label{subsec::models}
With appropriate choices of dark matter and gas profiles, it is now time to tie them together through hydrostatic equilibrium given as 
\begin{equation}
\boldsymbol{\nabla} P_g = -\rho_g \boldsymbol{\nabla} \Phi,
\label{eqn:clust:hydrostaticeq}
\end{equation}
where the gravitational potential, $\Phi$, can be found by solving Poisson's equation, $\nabla^2 \Phi = 4 \pi G \rho_{dm}$, which in an isotropic spherical geometry is given as
\begin{equation}
\frac{\partial \Phi}{\partial r} = \frac{4 \pi G}{r^2}\int_0^r \dif r' \rho_{dm}(r') r'^2.
\label{eqn:clust:poisson}
\end{equation}
Using the definition of the entropy as $S = p\rho^{-\gamma}$, where $\gamma$ is the adiabatic index, Eqn. [\ref{eqn:clust:hydrostaticeq}] can be rewritten as
\begin{equation}
\frac{\dif \rho}{\dif r} = - \frac{1}{\gamma}\frac{\partial \ln S}{\partial r}\rho - \frac{1}{\gamma S}\rho^{2-\gamma}\frac{\partial \Phi}{\partial r}.
\label{eqn:clust:model-int1}
\end{equation}
In order to simplify the subsequent derivations, we use the following dimensionless variables: $\rho^{*} \equiv \rho/\rho_0$, $y \equiv r/r_{dm}$, $x \equiv r/r_c = y r_{dm}/r_c$, where the subscript '0' refers to quantities at the cluster center.  Solving Poisson's equation for the gravitational acceleration, we obtain
\begin{equation}
\frac{\partial \Phi}{\partial r} = \frac{4\pi G \rho_{dm 0} r_{dm}}{y^2}\eta(y),
\label{eqn:clust:grav-acc}
\end{equation}
where the function $\eta(y)$ is defined as 
\begin{equation}
\eta(y) \equiv \mathrm{arcsinh}(y) - \frac{y}{\sqrt{1+y^2}}.
\label{eqn:clust:eta}
\end{equation}

Assuming the standard form for the pressure $P = \rho k_B T/\mu m_H$ where $\mu$ is the mean atomic number per nucleon (typically 0.6) and $m_H$ is the mass of the hydrogen atom, an equation for the density can be derived as
\begin{equation}
\frac{\dif \rho^*}{\dif y} = - \frac{\alpha}{\gamma}\frac{r_{dm}}{r_c}\frac{x^{\alpha -1}}{1 + x^{\alpha}}\rho^* - \frac{C}{1+x^{\alpha}}\frac{\rho^{* 2-\gamma}}{y^2}\eta(y),
\label{eqn:clust:model-final}
\end{equation}
where the coefficient $C$ encodes many of the physical parameters of the cluster
\begin{equation}
C \equiv \frac{4 \pi G \rho_{dm 0}\mu m_H r_{dm}^2}{\gamma k_B T_0}.
\label{eqn:clust:C-vars}
\end{equation}
In more physical units $C$ can be written as 
\begin{eqnarray}
C = &15.0&\left(\frac{\mu}{0.6}\right)\left(\frac{\rho_{dm 0}}{2.0\times 10^{-25}\,\mbox{g cm}^{-3}}\right)\\\nonumber
&\times&\left(\frac{r_{dm}}{500\,\mbox{kpc}}\right)^2\left(\frac{5/3}{\gamma}\right)\left(\frac{10\mbox{keV}}{k_B T_0}\right),
\label{eqn:clust:C-num}
\end{eqnarray}
which is dimensionless.  
The equation for $\rho$  can be formulated as either an ordinary differential equation or an integro-differential equation when the form of equation \ref{eqn:clust:poisson} is adopted for the gravitational acceleration.  The equation must be solved numerically to calculate a profile; however, it is singular at $y=0$.  Equation \ref{eqn:clust:model-final} can be integrated analytically in the limit of small $y$.  This integration yields a non-singular starting point for integration of
\begin{equation}
\rho^*(\epsilon) = 1 -\frac{1}{\gamma}\left(\frac{r_{dm}}{r_c}\right)^\alpha - \frac{C}{6} \epsilon^2.
\label{eqn:clust:taylor}
\end{equation}
With this initial condition, the atmosphere is integrated using a fourth-order Runge-Kutta solver based on Numerical Recipes \citep{press86}.  The fiducial parameters for our cluster are given in Table \ref{tab:clust:cluster-param}.  

\begin{figure}[htb!] 
\epsscale{0.45}
\centering
\includegraphics[clip=true, scale=0.44]{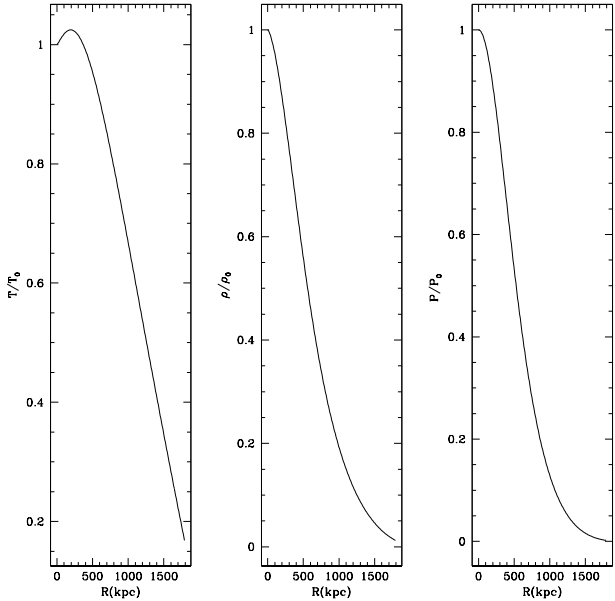}
\caption[The Fiducial Cluster Atmosphere Profiles]{The non-dimensional temperature, density, and pressure profiles are plotted for the fiducial cluster model given by Table \ref{tab:clust:cluster-param}.}\label{fig:clust:atm-model}
\end{figure}

\begin{deluxetable}{lcc}
\tablecolumns{3}
\tablecaption{Fiducial Cluster Parameters\label{tab:clust:cluster-param}}
\tablewidth{0pt}
\tablehead{
\colhead{Parameter Name} &
\colhead{Symbol} &
\colhead{Value} 
}
\startdata
Dark matter scale radius & $r_{dm}$ & 750 kpc \\
Entropy scale radius & $r_{c}$ & 1000 kpc \\
Central gas temperature & $T_0$ & 10 keV \\
Dark Matter Mass & $M_0$ & $5.6\times 10^{14}$ \Mo \\
Central dark matter density & $\rho_{dm 0}$ & $3.594\times 10^{-26}$ g cm$^{-3}$\\
Entropy profile exponent & $\alpha$ & 1.3 \\
King exponent & $n$ & 3\\
Constant for Eqn. [\ref{eqn:clust:model-final}] & C & 6.064\\
\enddata
\end{deluxetable}
Figure \ref{fig:clust:atm-model} shows the calculated non-dimensional profiles for the fiducial cluster model detailed in this section.  A nice property of this model is that the entropy is by construction increasing radially; thus, it is manifestly convectively stable by the Schwarschild criterion.  
\section{Method}\label{sec:method}
\subsection{Equations of MHD with Anisotropic Heat Conduction} \label{subsec:method:MHDeqn}
We solve the usual equations of ideal MHD with the addition of a vector heat flux, \boldmath$Q$\unboldmath, and a gravitational acceleration $\textbf{\em g} = -\boldsymbol{\nabla}\Phi$, 
\begin{equation}
\frac{\partial \rho}{\partial t} + \boldsymbol{\nabla}\cdot\left(\rho \boldsymbol{ v}\right) = 0,
\label{eqn:clust:MHD_continuity}
\end{equation}
\begin{equation}
\frac{\partial(\rho\boldsymbol{v})}{\partial t} + \boldsymbol{\nabla}\cdot\left[\rho\boldsymbol{vv}+\left(p+\frac{B^{2}}{8\pi}\right)\boldsymbol{I} -\frac{\boldsymbol{BB}}{4\pi}\right] + \rho\boldsymbol{\nabla}\Phi=0,
\label{eqn:clust:MHD_momentum}
\end{equation}
\begin{eqnarray}
\frac{\partial E}{\partial t} &+& \boldsymbol{\nabla}\cdot\left[\boldsymbol{v}\left(E+p+\frac{B^{2}}{8\pi}\right) - \frac{\boldsymbol{B}\left(\boldsymbol{B}\cdot\boldsymbol{v}\right)}{4\pi}\right] \\ \nonumber
&+&\boldsymbol{\nabla}\cdot\boldsymbol{Q} +\rho\boldsymbol{\nabla}\Phi\cdot\boldsymbol{v}=0,
\label{eqn:clust:MHD_energy}
\end{eqnarray}
\begin{equation}
\frac{\partial\boldsymbol{B}}{\partial t} - \boldsymbol{\nabla}\times\left(\boldsymbol{v}\times\boldsymbol{B}\right)=0,
\label{eqn:clust:MHD_induction}
\end{equation}
where the symbols have their usual meaning. The total energy $E$ is given as
\begin{equation}
E=\epsilon+\rho\frac{\boldsymbol{v}\cdot\boldsymbol{v}}{2} + \frac{\boldsymbol{B}\cdot\boldsymbol{B}}{8\pi},
\label{eqn:clust:MHD_Edef}
\end{equation}
and the internal energy, $\epsilon=p/(\gamma-1)$.  We assume
$\gamma=5/3$ throughout.

The heat flux, $\boldsymbol{Q}  = \boldsymbol{Q}_C + \boldsymbol{Q}_{iso}$, 
where $\boldsymbol{Q}_C$, the anisotropic term is due to Coulombic conduction parallel to magnetic field lines.  The isotropic heat flux is given solely for comparison to the effect of anisotropic conductivity.  The anisotropic conductivity is either set to a constant thermal diffusivity or given as the expression
\begin{equation}
\boldsymbol{Q}_{C} = - n k_B \chi_{C}(T, n) \boldsymbol{\hat{b}\hat{b}}\cdot\boldsymbol{\nabla}T,
\label{eqn:clust:coulombic}
\end{equation}
where the diffusivity is given by the Spitzer conductivity \citep{spitz62} and $\boldsymbol{\hat{b}}$ is a unit vector in the direction of the magnetic field.  The Spitzer thermal diffusivity is parameterized appropriately for the ICM as
\begin{equation}
\chi_{C}(T,n) = 8\times 10^{31}\left(\frac{T}{10\,\mbox{keV}}\right)^{5/2}
\left(\frac{n}{5\times 10^{-3}}\right)^{-1} \;\mbox{cm}^2\mbox{s}^{-1}.
\label{eqn:clust:conductivity}
\end{equation}
Note that this is a \textit{thermal diffusivity}, and it is inversely dependent on the density.  When one speaks of the Spitzer conductivity, one often is speaking of $\kappa_{Sp} = n k_B \chi_C$ which has only the well-known $T^{5/2}$ dependence and no density dependence.  
\subsection{Initial Conditions} \label{subsec:method:IC}
The initial conditions of the intracluster medium are described in detail in previous sections and the resulting plasma parameters are presented visually in Figure \ref{fig:clust:atm-model}.  A database from the Runge-Kutta solver is linearly interpolated to fill all space.  The dark matter potential given by Equation \ref{eqn:clust:grav-acc} is applied as a static gravitational acceleration on the grid; thus, achieving hydrostatic equilibrium.

We utilize two different magnetic field configurations in this study.  The first initial magnetic field geometry is chosen to be both simple and compelling.  Namely, the magnetic field is initialized to be purely azimuthal in the $x$--$y$ plane at all radii such that $r\le r_{max}$.  Essentially, this configuration is a sphere of nested circular magnetic field loops of constant strength.  Numerically this configuration is realized by utilizing the vector potential, where $\boldsymbol{B} = \boldsymbol{\nabla}\times \boldsymbol{A}$.  For a constant magnetic field of strength $B_0$ defined as described, the vector potential is given as
\begin{equation}
\boldsymbol{A} = - B_0 \sqrt{x^2 + y^2} \;\hat{\boldsymbol{z}}.
\label{eqn:clust:Az}
\end{equation}
The vector potential is calculated at cell corners and then differenced to calculate the face-centered magnetic fields.  The advantage of this magnetic field geometry is that the heat flux is zero in the initial configuration; thus, all subsequent evolution is due to the MTI.  The fiducial magnetic field strength is chosen to be 1 nG, possibly moderately higher than the poorly constrained primordial field but substantially less than today's observed field.

A second magnetic field geometry is chosen to be a more realistic representation for the tangled fields observed in clusters today.  There is little theoretical guidance for the magnetic field power spectrum in the ICM, so we choose a Kolmogorov power spectrum for the magnetic field, initialized in Fourier space as
\begin{equation}
\tilde{A}(k) = \tilde{A}_0\left(\frac{k}{k_\textrm{peak}}\right)^{-\alpha},
\label{eqn:A-k}
\end{equation}
where $k_{\textrm{peak}}$ is chosen as the wavenumber corresponding to 2--4 times the grid scale.  We utilize the Fast Fourier Transform (FFT) as was done in \citet{rs99} to calculate the vector potentials in real space.  We also randomize the phase to avoid correlating the modes.  The magnetic energy scales as $B^2 \propto k^{-2(\alpha-1)}$.  We initialize each Gaussian component separately, so for the componentwise Kolmogorov spectrum, the appropriate choice is $\alpha= -17/6$ to give the familiar scaling for energy of $k^{-11/3}$.  This $k$-space scaling for Kolmogorov turbulence is appropriate for initializing the 1D power spectrum (as opposed to the 3D power spectrum) in each direction separately.  We numerically check that the divergence of the magnetic field remains zero and renormalize the magnetic energy such that $\langle B^2\rangle = \langle B_0^2\rangle$.  

\subsection{Timescales} \label{subsec:method:timescale}
In this section we discuss several of the characteristic timescales for a cluster as was outlined in \citet{ps07b}. The Brunt-V\"ais\"al\"a frequency can be written in a more useful form for the cluster problem as 
\begin{equation}
N^2 = \frac{1}{\gamma}\frac{\partial \Phi}{\partial r}\frac{\partial \ln S}{\partial r}.
\label{eqn:clust:N2}
\end{equation}
The maximum growth rate is given by the isothermal limit of the Brunt-V\"ais\"al\"a frequency, namely
\begin{equation}
\sigma^2_{max} = \frac{\partial \Phi}{\partial r}\frac{\partial \ln T}{\partial r}.
\label{eqn:clust:sigmamax}
\end{equation}
Table \ref{tab:clust:timescales} examines these timescales in the galaxy cluster at a fiducial radius of 450 kpc.  Several of these timescales are dependent on the length scale of interest.  Thus, they are considered on global scales, roughly a wavelength of 1 Mpc, and the scale of the mean free path, roughly 30 kpc.  The magnetic field is assumed to be 1 nG for these calculations.  
\begin{deluxetable*}{lccc}
\tablecolumns{4}
\tablecaption{Timescales of the ICM \label{tab:clust:timescales}}
\tablewidth{0pt}
\tablehead{
\colhead{Timescale} &
\colhead{Definition} &
\colhead{$\lambda=$ 1 Mpc (Myrs)} &
\colhead{$\lambda=$ 30 kpc (Myrs)} 
}
\startdata
Sound crossing time & $\tau_s = \frac{\lambda}{c_s}$ & 
$610$ & $18$\\
Alfv\'{e}n crossing time & $\tau_A = \frac{\lambda}{v_A}$ &
$5.4\times 10^{7}$ & $1.6 \times 10^{6}$ \\
Thermal conduction time & $\tau_{\chi} = \frac{\lambda^2}{\chi_C}$ &
$2.3\times 10^3$ & $2.1$\\
Brunt-V\"ais\"al\"a time & $\tau_{N} = N^{-1}$ &
$810$ & $810$\\
Growth time & $\tau_{\sigma} = \sigma_{max}^{-1}$ &
$880$ & $880$\\
\enddata
\end{deluxetable*}

As can be seen by examination, the MTI has tens of growth times during a Hubble time, allowing significant rearrangement of the atmosphere.  The magnetic field estimated using primordial values clearly plays no role, even on the length scale of the mean free path.   

On mean free path timescales, the conduction time is faster than the sound crossing time. The potential worry is due to the thought that the heat flux is saturated in the sense of \citet{cm77}.  The right comparison there is not the conduction time to the mode sound crossing time but the electron thermal velocity to the velocity needed for electrons to transport the heat.  For the atmosphere considered here, the saturated heat flux is several orders of magnitude larger than the actual heat flux.  
\subsection{Numerical Methods} \label{subsec:method:numerical}
We use the 3D version of the Athena MHD code \citep{gs08,sg08} to simulate the MTI far into the nonlinear regime. Both the MHD and the heat transport methodology and tests have been described in these references and \citet{ps05}, respectively.  The thermal diffusivity is either set as a constant or permitted to vary spatially according to the standard Spitzer prescription.  Here we adopt the adopt the method of \citet{sh07} for handling variable thermal conductivities.  Namely, the product of density and thermal diffusivity is interpolated to the cell faces using a harmonic average, \textit{e.g.} in one dimension as
\begin{equation}
\frac{2}{(n\chi)_{i+1/2}} = \frac{1}{(n\chi)_i} + \frac{1}{(n\chi)_{i+1}}.
\label{eqn:clust:conductivity-var}
\end{equation}
This harmonic averaging prevents the Courant condition from becoming severe due to discontinuous diffusivities and densities at interfaces, should they develop.  The timestep is determined with respect to the maximum thermal diffusivity on the grid and the conduction module is sub-cycled.  

All the runs described in this chapter are performed on a uniform Cartesian three-dimensional grid with the initial cluster atmosphere described in \S\ref{sec:model}.  The domain extends from $-3.0\times 10^{24}$ cm to $+3.0\times 10^{24}$ cm (968 kpc) in each direction.  Most of the runs presented in this paper are simulated at a resolution of $(128)^3$ or corresponding to a zone size of 15 kpc respectively.  The simulations are typically run for $3\times 10^{17}$ s or 9.5 Gyr, a substantial fraction of cosmic time.  The magnetic field is initialized in a region $r_{max}$ which is slightly less than the length of the cube.  Typically for our simulations $r_{\textrm{max}} = 2.5\times 10^{24}$ (806 kpc).  Thus, the initial magnetic field is confined to a spherical region.  Since there is no magnetic field outside of $r_{\textrm{max}}$, there is initially no conduction in this region. 

We use modified reflecting boundary conditions for all the MHD variables, in which the pressure and density are extrapolated in the ghost zones. In order to prevent any negative values for pressure or density in the ghost zones, we introduce pressure and density floors applied only in the ghost zones. The magnetic field and velocity field components are simply reflected at the boundary.  The heat fluxes are forced to be adiabatic by setting \mbox{$\boldsymbol{Q}\cdot\boldsymbol{\hat{N}}=0$} on the boundary.  

To seed multiple modes of the MTI and break symmetry,we add Gaussian white noise perturbations to the velocity field such that the applied perturbation is a fixed fraction of the local sound speed everywhere within the cluster.  A typical perturbation is 1\% of the sound sound speed in local velocity magnitude.  In the absence of applied perturbations, it is reassuring to note excellent Cartesian symmetry is maintained.  

\subsection{ICM Atmosphere Stability for MHD-only Runs} \label{subsec:method:stability}
In order to both show that these numerical methods are valid and to provide a basis for comparison, we introduce a reference calculation in which the atmosphere is initialized as previously described, except thermal conduction is disabled.  In this case, the MTI is not active and the stability properties of the background atmosphere can be deduced.  

The background state is run for for a time of $3\times 10^{17}$ s (9.6 Gyr) on a $(64)^3$ grid, at which point it is analyzed.  The cluster remains relatively quiescent.  Of course, there is a slight mismatch resulting from discretizing a spherical problem on a Cartesian grid. The maximum Mach number reached anywhere on the grid is $9.8\times 10^{-3}$ and the rms Mach number is even smaller.  Another characterization of the quiescence is the change in the central pressure.  Over the entire run time the maximum change in the central pressure from the initial condition is only 0.2\%.  It is also useful to look at the conservation of energy. It is not just internal energy that must be considered, but also the total gravitational binding energy of the cluster:
\begin{equation}
\frac{\dif}{\dif t}\left<e + \frac{1}{2}\rho\boldsymbol{v}\cdot\boldsymbol{v} + \frac{1}{2}\boldsymbol{B}\cdot\boldsymbol{B} + \rho\Phi\right> = 0,
\label{eqn:clust:conservation-E}
\end{equation}
where the angle brackets reflect a volume average.  For the case where the MTI is active, the density profile is modified, and the last term is important to consider.  By this measure, the total energy is conserved to better than 0.1\% for both the MHD-only and MHD plus conduction cases.  

Finally, it is useful to introduce a diagnostic at this point which reflects the net alignment of the magnetic field with the radial direction.  The diagnostic is mathematically represented as \bdotr, which is the overlap of the magnetic field unit vector and the radial unit vector, averaged over the cluster region in which the field was originally initialized.  The original azimuthal field construction would have this quantity be ideally zero, but discretization errors give an initial value of $\langle |\boldsymbol{\hat{b}}\cdot \boldsymbol{\hat{r}}|\rangle_0 = 2.3\times 10^{-4}$ for $(64)^3$.  At the conclusion of the MHD only run, this value has risen to $0.02$, a small change to be compared to the runs with conduction later.  In addition, in the MHD-only run, there is approximately a 5\% decay in magnetic field energy through reconnection, which is also the primary source of the change of the magnetic field orientation.  Thus, in this section, we have characterized the evolution of the cluster atmosphere with MHD-only runs, and shown that this atmosphere remains very quiescent.
\begin{deluxetable*}{lccccc}
\tablecolumns{6}
\tablecaption{Initial Properties of Nonlinear Runs  \label{tab:clust:runs}}
\tablewidth{0pt}
\tablehead{
\colhead{Run} &
\colhead{Conduction Type} &
\colhead{Conductivity ($\textrm{cm}^2\textrm{s}^{-1}$}) &
\colhead{Field Geometry} &
\colhead{$B_0/\left(4\pi\right)^{1/2}$ ($\mu$G)} &
\colhead{$Resolution$} 
}
\startdata
F0........... & None & 0 & Circular & $10^{-9}$ & $64^3$\\
A1........... & Anisotropic & $8\times 10^{31}$ & Circular & $10^{-9}$ & $128^3$\\
A2........... & Anisotropic & Spitzer & Circular & $10^{-9}$ & $128^3$\\
A3........... & Anisotropic & Spitzer & Tangled & $10^{-9}$ & $128^3$\\
A4........... & Anisotropic & Spitzer & Circular & $10^{-9}$ & $64^3$\\
A5........... & Anisotropic & Spitzer & Circular & $10^{-9}$ & $256^3$\\
A6........... & Anisotropic & Spitzer & Tangled & $10^{-9}$ & $64^3$\\
A7........... & Anisotropic & Spitzer & Tangled & $10^{-9}$ & $256^3$\\
A8........... & Anisotropic & Spitzer & Circular & $10^{-6}$ & $128^3$\\
I1........... & Isotropic & $8.0\times 10^{30}\,T^{5/2}n^{-1}$ & Circular & $10^{-9}$ & $128^3$\\
I2........... & Isotropic & $2.4\times 10^{31}\,T^{5/2}n^{-1}$ & Circular & $10^{-9}$ & $128^3$\\
\enddata
\end{deluxetable*}
\section{Simulations of the MTI}\label{sec:sim}
In order to simulate the effects of the MTI on the intracluster medium, We have performed a number of simulations that are listed in Table \ref{tab:clust:runs}.  The key runs of interest are the ones with purely anisotropic conductivity labeled A1--A8.  Run A1 has a fixed thermal diffusivity and provides the cleanest physics laboratory for understanding the MTI. Runs A2--A7 have a variable Spitzer thermal diffusivity.  These runs also constitute a resolution and geometry study.  Run A8 is the same as run A2, except the magnetic field is three orders of magnitude larger.  For variable conductivity, the global conduction timestep is limited by the largest conductivity anywhere on the grid, which has made $(256)^3$ achievable only with several tens of thousands of processor hours.

We also provide a number of runs for comparison purposes.  First, the fiducial run we have already outlined in \S\ref{subsec:method:stability} is labeled F0.  This run shows what the background state is in the absence of any thermal conduction.  We also provide runs I1--I2 which utilize purely isotropic conductivity at a fixed fraction of the Spitzer conductivity to highlight the difference between the instability and simple isotropic conductions. 
\subsection{Fixed Anisotropic Thermal Conduction}\label{subsec:sim:fixed-ani}
\begin{figure}[htb!] 
\epsscale{0.50}
\centering
\includegraphics[clip=true, scale=0.45]{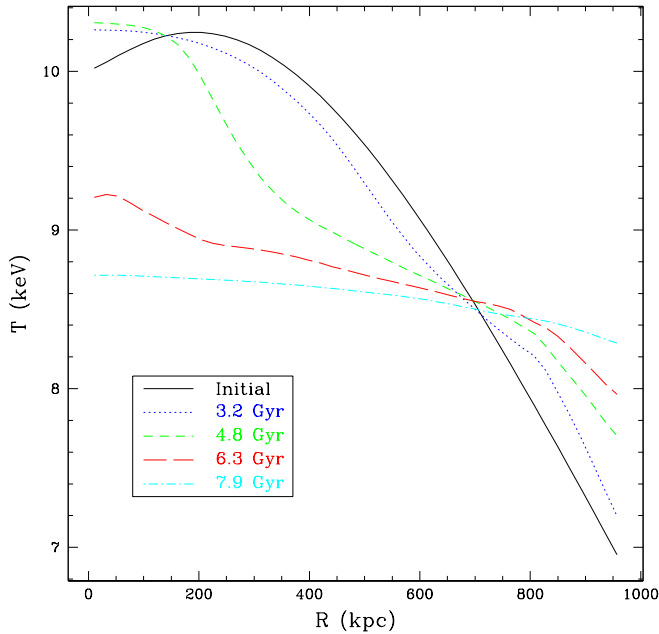}
\caption[Temperature Profiles for the MTI for Run A1]{The MTI dramatically rearranges the radial temperature profiles for Run A1. }\label{fig:clust:A1-temp}
\end{figure}

We begin our discussion with the simplest illustrative case of this paper, Run A1.  This simulation of the MTI was performed with a fixed thermal diffusivity of $8.0\times 10^{31}$ cm$^2$s$^{-1}$.  In addition, the magnetic geometry is initialized with simple circular magnetic field lines.  Figure \ref{fig:clust:A1-temp}
shows the dramatic effect of the MTI on the radial shell-averaged temperature profiles.  In a time short compared to the Hubble time, the temperature profile is quite dramatically rearranged.  By 7.9 Gyr, the initial temperature profile is almost completely erased, and the cluster driven towards isothermality.  Concomitantly the density in the central regions is driven somewhat higher with the central density peaking at $6.6\times 10^{-27} \;\mbox{g}\,\mbox{cm}^{-3}$.  While initially some heat is transported inwards, Brehmstrahlung cooling scales like $n_e^2$; thus, with the addition of cooling, this density increase may enhance the cooling flow problem.  

\begin{deluxetable*}{lcccccc}
\tablecolumns{7}
\tablecaption{Saturated Properties of Nonlinear Runs  \label{tab:satprop}}
\tablewidth{0pt}
\tablehead{
\colhead{Run} &
\colhead{$\delta\langle B^2\rangle$} &
\colhead{RMS Mach\tablenotemark{a}} &
\colhead{Max. Mach\tablenotemark{a}} &
\colhead{$\langle\boldsymbol{\hat{b}}\cdot\boldsymbol{\hat{r}}\rangle$} &
\colhead{$f_{\mathrm{Spitz}}$\tablenotemark{b}} &
\colhead{$t_{\mathrm{max. KE}}$ (Gyr)}
}
\startdata
F0........... & 0.95 & $2.4\times 10^{-3}$ & $9.8\times 10^{-3}$ & 
0.05 & ... & 0 \\
A1........... & 18.4 & 0.03 & 0.3 & 0.62 & 0.28 & 7.7\\
A2........... & 19.0 & 0.04 & 0.3 & 0.59 & 0.28 & 7.0\\
A3........... & 17.2 & 0.03 & 0.1 & 0.67 & 0.45 & 1.8\\
A4........... & 13.2 & 0.04 & 0.3 & 0.60 & 0.43 & 5.2\\
A5........... & 23.5 & 0.04 & 0.3 & 0.59 & 0.26 & 7.1\\
A6........... & 14.6 & 0.05 & 0.2 & 0.70 & 0.49 & 1.7\\
A7........... & 59.8 & 0.05 & 0.3 & 0.68 & 0.43 & 1.8\\
A8........... & 0.95 & 0.04 & 0.2 & 0.44 & 0.13 & 7.7\\
I1........... & 1.1  & $1.0\times 10^{-3}$ & $1.2\times 10^{-3}$ & 
0.19 & ... & 1.0 \\
I2........... & 1.1  & $1.1\times 10^{-3}$ & $1.4\times 10^{-3}$ & 
0.19 & ... & 0.9 
\enddata
\tablenotetext{a}{Evaluated at the peak of kinetic energy.}
\tablenotetext{b}{Evaluated at the peak of \bdotr, normalized to the instantaneous Spitzer heat flux.}
\end{deluxetable*}
Simultaneously, the MTI is driving a magnetic dynamo which acts to amplify the magnetic field.  This is quantified as
\begin{equation}
\delta\langle B^2 \rangle \equiv \frac{\langle B^2 \rangle_{\textrm{fin}}}{\langle B^2 \rangle_{\textrm{init}}},
\label{eqn:clust:b-amp}
\end{equation}
where the angle brackets denote a volume average over the sphere with $r<r_{\textrm{max}}$.  For Run A1, the magnetic energy density is amplified by a factor of 18.4  from the initial energy density. There is actually somewhat more magnetic dynamo action, as some of the amplified flux is transported by penetrative convection to $r>r_{\textrm{max}}$ and is not included in this average amplification factor.   The saturation properties of this run and all other simulations are given in Table \ref{tab:satprop}.  This magnetic field increase is significant, but clearly not entirely sufficient to raise the primordial field to its present-day value.  Certainly, much of the increase comes from flux-freezing during gravitational collapse, which is not included here.  In addition, smaller-scale field increases could come from purely kinetic processes such as the Alfv\'{e}n wave cascade or other kinetic process \citep[e.g.,][]{sc06a, sc06b}. The large scale turbulence can serve as energy input for this process.

\subsection{Spitzer Anisotropic Conduction}\label{subsec:varcond}
\begin{figure}[htb!] 
\epsscale{0.50}
\centering
\includegraphics[clip=true, scale=0.45]{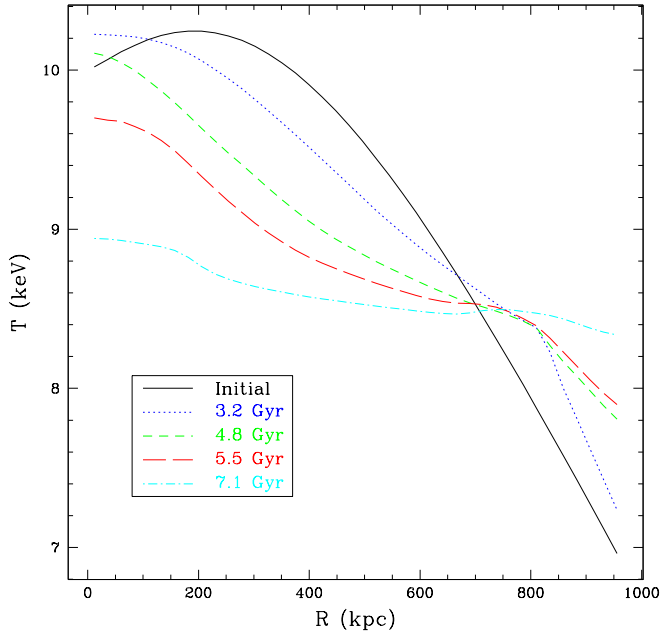}
\caption[Temperature Profiles for the MTI for Run A2]{The shell-averaged radial temperature for circular magnetic fields with Spitzer conductivity in run A2 evolves significantly. }\label{fig:clust:A2-temp}
\end{figure}

We proceed to discuss the more realistic fiducial case of run A2.  Here, instead of a fixed conductivity as in A1, we utilize the full variable Spitzer conductivity as given in Eqn. [\ref{eqn:clust:conductivity}].  We begin by examining the thermal evolution of A2 in Figure \ref{fig:clust:A2-temp}.
The evolution of the temperature profile is somewhat smoother than in the case with fixed conductivity and has approached a fairly isothermal cluster by about 7 Gyr after the start of the run.  Just as a reminder, the anisotropic conduction is initially zero due to $B=0$ beyond $r_{\textrm{max}}$, which is slightly greater than 800 kpc.

\begin{figure}[htb!] 
\epsscale{0.50}
\centering
\includegraphics[clip=true, scale=0.45]{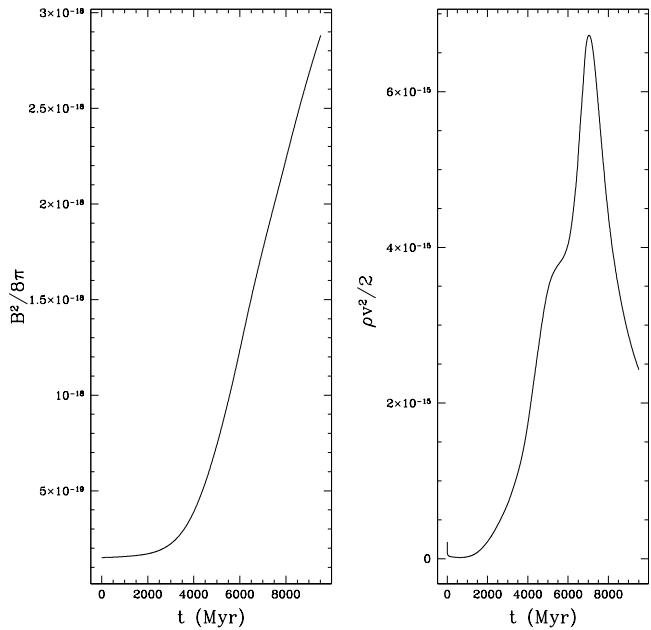}
\caption{The volume-averaged evolution of the magnetic and kinetic energy during the course of run A2 show magnetic field amplification and significant motions. }\label{fig:clust:A2-time}
\end{figure}

Figure \ref{fig:clust:A2-time} shows the time evolution of the volume-averaged magnetic and kinetic energy for run A2.  The magnetic energy is continually amplified during the entire evolution, reaching a peak amplification of $\delta\langle B^2\rangle\approx 19.0$.  The kinetic energy reaches a peak around 7.7 Gyr.  At this time, the RMS Mach number, where $\textrm{Mach}\equiv|\boldsymbol{v}|/c_s$, averaged over the cluster volume is 0.04.  The maximum Mach number of the turbulence at any given zone in the cluster is 0.3.
The kinetic energy decays somewhat as the source of free energy, the temperature gradient, has been mostly exhausted.  In practice, clusters are constantly being heated by gravitational infall and AGN activity in the cluster core.   These heating terms are neglected but clearly would contribute to maintaining the temperature gradient and source of free energy; thus, the decay of turbulence is likely overpredicted in the simulation.

\begin{figure}[htb!] 
\epsscale{0.50}
\centering
\includegraphics[clip=true, scale=0.45]{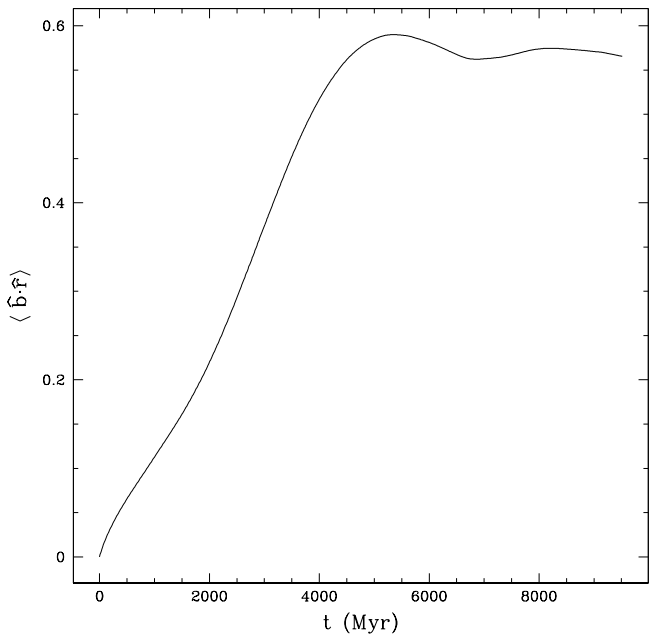}
\caption{The volume-averaged radial component of the magnetic field, \bdotr, is amplified significantly during the course of run A2.}\label{fig:A2-bdotr}
\end{figure}

Perhaps the most interesting part of the evolution dynamics is that the magnetic field geometry is significantly rearranged by the MTI.  Initially, the magnetic field geometry was almost perfectly azimuthal, lying completely in the $x$--$y$ plane.  At the conclusion of the simulation, the magnetic field is highly biased in the radial direction, with $\langle|\boldsymbol{\hat{b}}\cdot\boldsymbol{\hat{r}}|\rangle = 0.62$ at maximum.  Figure \ref{fig:A2-bdotr} shows the evolution of this metric.  The tendency of the MTI to align the magnetic field with the background temperature gradient was previously seen in the 3D stratified box simulations presented in \citet{ps07a}.  This large scale rearrangment of the magnetic field has potential observational consequences which will be discussed later.  
 
It is notable to compare this simulation to the fiducial simulation, F0, in which conduction does not play a role.  In the absence of a dynamo driven by conduction, the magnetic field actually decays by reconnection.  In F0, the flow remains very quiescent and the magnetic field almost completely retains its initial azimuthal geometry.  

The same simulation as run A2 has also been performed at $(64)^3$ (run A4) to provide a resolution study.  In general, as is to be expected, the prime difference is the degree of amplification of the magnetic field.  Small scale wavelengths grow fastest for the MTI; thus, higher resolution permits more small-scale field growth at somewhat faster timescales.  The higher resolution also permits greater tangling of the field by convective motions.  The Mach number is slightly higher in lower resolution runs; whereas the magnetic field geometry remains fairly constant.  Given the dyanmically weak magnetic fields achieved after 9.5 Gyr of growth, we are not yet in the limit where magnetic tension is playing a significant role on the small scales.  Thus, the magnetic field amplification is not yet saturated with the tangled field amplification resulting in $\delta\langle B^2\rangle\approx 59.8$ at $256^3$ and 19.0 at $128^3$.  Higher resolution, while not computational tractable, will likely yield even higher magnetic field amplification.  Aside from this exception, we are not seeing any significant trends with resolution.
\subsection{Heat Flux Diagnostics}\label{subsec:heatflux}
\begin{figure}[htb!] 
\epsscale{0.50}
\centering
\includegraphics[clip=true, scale=0.45]{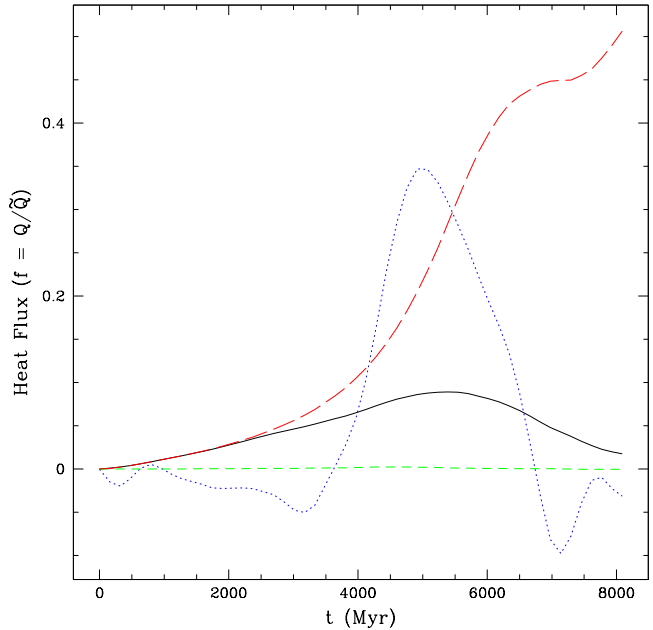}
\caption{Here we show the evolution of the radial heat fluxes for run A2 in a 100 kpc shell located 450 kpc from the cluster center.  The conductive heat flux normalized to the \textit{initial} fiducial heat flux (\textit{black, solid line}) reaches a maximum before decreasing, while normalized to the \textit{instantaneous} fiducial heat flux (\textit{red, long-dashed line}) continues to increase.  The heat fluxes due to mass advection (\textit{blue, dotted line}) and convection (\textit{green, short-dashed line}) are shown normalized to the initial fiducial heat flux.} \label{fig:A2-heatflux}
\end{figure}

We also analyze the cluster heat flux from our stored time slices at the conclusion of a run.  We will focus on the radial component of the heat fluxes in a spherical shell of width 100 kpc, centered on a radius of 450 kpc from the cluster center.  We begin by defining a fiducial conductivity to be the Spitzer conductivity in this shell if the conduction were isotropic at the Spitzer value, namely:
\begin{equation}
\widetilde{Q}_{r} = - n k_B \chi_C(T,n) \frac{\dif T}{\dif r}.
\label{eqn:Qfid}
\end{equation}
This value is the same as for anisotropic thermal conduction with purely radial field lines.  In Figure \ref{fig:A2-heatflux}, we compare the heat flux due to conduction to this Spitzer value.  However, there remains some ambiguity as the temperature profile evolves over the course of the simulation.  Thus, we normalize the conductive heat flux to both the \textit{initial} Spitzer heat flux and the \textit{instantaneous} Spitzer heat flux.  We define the Spitzer fraction, the oft-quoted \textit{f} paramater as 
\begin{equation}
f_{\textrm{Spitz}} = Q_{\textrm{cond}}/\widetilde{Q}.
\label{eqn:fspitz}
\end{equation}
In Table \ref{tab:satprop}, we utilize the instantaneous fiducial heat flux at the time of maximum \bdotr for normalization of this quantity.  Figure \ref{fig:A2-heatflux} shows that as the ICM temperature gradient is depressed, the conductive heat flux normalized to the instantaneous heat flux increases substantially, until $f_\textrm{Spitz}$ is larger than 0.4. 

Meanwhile, calculating the heat flux due to the transport of enthalpy is worth a careful treatment.  The advection of enthalpy can be written as
\begin{equation}
\boldsymbol{Q}_{\textrm{enthalpy}} = \left(\frac{\gamma}{\gamma -1}\right) p\boldsymbol{v},
\label{eqn:qenthalpy}
\end{equation}
where we neglect the transport of kinetic and magnetic energy since the flows are both low mach number and high-$\beta$, respectively.  We can split this heat flux into an advective term proportional to a mass flux ($\propto n\boldsymbol{v}$) and a term that is a purely convective heat flux.  Writing the pressure in Eqn.[\ref{eqn:qenthalpy}] as $nk_B T$, we define several quantities:
\begin{eqnarray}
n &=& \langle n\rangle + \delta n\\
v &=& \langle v\rangle + \delta v\\
T &=& \langle T\rangle + \delta T,
\end{eqnarray}
where, for example $\langle n\rangle$ is the spatial average of the number density in the shell and $\delta n$ is the local deviation of a quantity from its average.  In this formulation the new total heat flux becomes
\begin{eqnarray}
Q_r &=& \frac{\gamma}{\gamma-1}k_B\langle nTv_r\rangle\\ \nonumber
 &=& \frac{\gamma}{\gamma-1}k_B\Bigl\langle(\langle n\rangle + \delta n)(\langle T\rangle + \delta T)(\langle v_r\rangle + \delta v_r)\Bigl\rangle.
\end{eqnarray}
In future equations, we drop the $r$ subscript.   In simplifying this expression, terms that are proportional to the average of a delta quantity, like $\langle \delta n\rangle$ vanish by definition.  We can then separate it into two quantities: one due to a mass flux, identified as terms that scale as $nv$, and the remaining terms into a convective flux.  Thus,
\begin{eqnarray}
\label{eqn:qmassadv}
Q_{adv} &=& \frac{\gamma}{\gamma-1}k_B\left(\langle n\rangle\langle T\rangle\langle v\rangle +\langle T\rangle\langle\delta n\delta v\rangle\right),\\
\label{eqn:qconv}
Q_{conv}  &=& \frac{\gamma}{\gamma-1}k_B\left(
  \langle v\rangle\langle\delta n\delta T\rangle +
  \langle n\rangle\langle\delta v\delta T\rangle\right. \\ \nonumber
&+& \left.\langle\delta n\delta T\delta v\rangle\right),
\end{eqnarray}
where the second term of equation (\ref{eqn:qconv}) dominates the convective flux.

Figure \ref{fig:A2-heatflux} shows these two heat fluxes normalized to the initial fiducial heat flux.  The mass advection heat flux is enhanced due to the inability of the dark matter profile to adjust to changes in the ICM structure, so it is not a physically important quantity.  What is interesting, however, is that the convective heat flux is quite small.  In fact, it is roughly four orders of magnitude smaller than the conductive heat flux.  The small convective heat flux seems to result from two principal causes.  First, unlike in solar convection, there is a second channel open for energy flow, namely directly through conduction which is quite efficient.  Second, as the magnetic fields become more radial, the buoyant driving is significantly reduced, reducing the baseline convective motions.  

At this point it is also appropriate to discuss run A8.  Run A8 is chosen to be initialized at the MTI stability boundary for strong magnetic fields, namely $kv_A \sim \sigma_{\textrm{max}}$, for wavenumbers corresponding to scales slightly smaller than the cluster radius.  In this limit, all but the largest scale modes are suppressed entirely by magnetic tension.  There is both theoretical motivation and numerical evidence that the conductive flux increases as the initial magnetic field increases.  
We do indeed find that for run A8, the convective flux is within an order of magnitude of the conductive flux; however, this phase lasts only a very short time (less than 500 Myr) and the net convective flux never is larger than $0.035 \,\widetilde{Q}$.  In addition, for this run, the total heat flux carried is considerably smaller than the other runs.  Thus, we conclude that conduction is the dominant heat transport mechanism and that convective heat transport can never be larger than a small fraction of the Spitzer conductivity for MTI-driven turbulence in galaxy clusters.  
\subsection{Comparison to Isotropic Conductivity}\label{subsec:sim:fixed-iso}
In order to emphasize the nature of the MTI as a truly convective instability, we have performed simulations with purely isotropic conductivity.  A long-favored approach to understanding thermal conduction in galaxy clusters has been to assume a tangled magnetic field geometry passively determined by hydrodynamic turbulence.  This tangled geometry then provides an effective thermal conductivity that is a specified fraction of Spitzer \citep{rr78, cc98, nm01}.  A major point of this paper is that such an approach is incorrect since thermal conduction along the magnetic field self-consistently modifies the magnetic field geometry, resulting in an $f_{\textrm{Spitzer}}$ that evolves with time and is not consistent with these theories.  Thus, here we present simulations I1 and I2 which have a thermal diffusivity fixed at 10\% or 30\% of the Spitzer value, respectively, for comparison purposes.  

The temperature profiles of these two runs evolve in ways quite similar to that of run A2 and A3, reaching fairly isothermal in 5-7 Gyr. A much more illustrative difference is found by considering the time evolution of the two runs.  The runs are initialized with a Mach 0.01 velocity perturbation that decays in time quite quickly, peaking at 1 Gyr or less, as opposed to 7 Gyr for run A2.  Without the MTI, the magnetic field is barely amplified at all, $\delta\langle B^2\rangle \approx 1.3$ for both cases. 

Finally, the magnetic field geometry is somewhat rearranged by the cluster motions, but does not even reach a geometrically isotropic state, with \bdotr\, reaching a maximum of 0.19.  Of course, we started in a state of pure azimuthal fields, so the key message here is that a field that started isotropically distributed would remain isotropically distributed---there is no force here pushing the field to be radially biased as with the MTI.  Clearly, the comparison between these isotropic conduction calculations and the preceeding anisotropic heat conduction calculation shows that the MTI plays a significant role in setting a self-consistent solution for magnetic field amplification, convective velocities, and for the magnetic field geometry of a cluster. 
\subsection{Tangled Magnetic Fields}\label{subsec:tangled}
\begin{figure}[htb!] 
\epsscale{0.50}
\centering
\includegraphics[clip=true, scale=0.45]{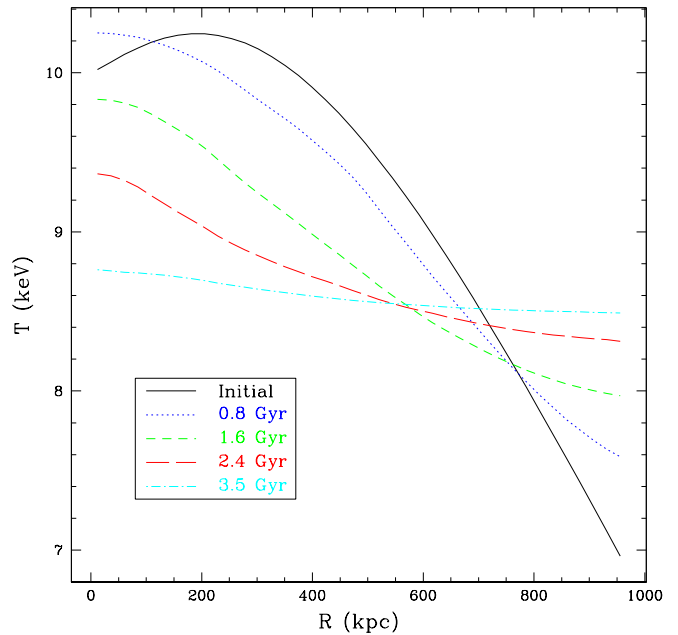}
\caption{The evolution of the shell-averaged radial temperature for tangled magnetic fields with Spitzer conductivity in run A3 is quite rapid. Note that the times plotted are roughly a factor of two earlier than for run A2 in Figure \ref{fig:clust:A2-temp}.}\label{fig:A3-temp}
\end{figure}

\begin{figure}[htb!] 
\epsscale{0.50}
\centering
\includegraphics[clip=true, scale=0.45]{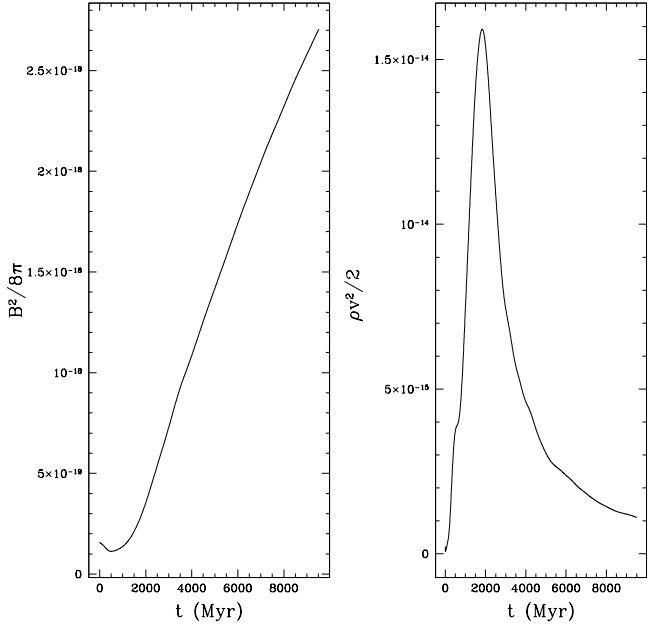}
\caption{The volume-averaged evolution of the magnetic and kinetic energy during the course of run A3. }\label{fig:A3-time}
\end{figure}

We now turn to the case of tangled magnetic fields, examined in run A3.  This run has the exact same set-up as the previously discussed runs, except the intial magnetic field is a turbulent magnetic field distributed according to the Kolmogorov scaling law.  Thus, in this initial case, there is a zeroth order anisotropic heat flux that exists independent of the MTI's evolution.  Figure \ref{fig:A3-temp}
shows the fairly rapid evolution of the temperature profile for the tangled field, reflecting this zeroth order radial heat flux. The cluster reaches a fairly isothermal temperature in 3.5 Gyr.  While the temperature gradient is being suppressed, the magnetic energy is amplified by a modest factor of 17.2, excluding the flux pumped to radii larger than $r_{\textrm{max}}$. 
Figure \ref{fig:A3-time} shows the evolution of the magnetic and kinetic energy over the run.
The kinetic energy peaks at 1.8 Gyr, much earlier and slightly higher in amplitude than the circular field case.  The short wavelength-modes that exist in the tangled field geometry grow and saturate on somewhat faster timescales than the largest modes in the problem.

An interesting property of the resultant turbulence is seen in examining Figure \ref{fig:A3-time}.  The kinetic energy at its peak is significantly larger than the magnetic energy.  At the peak of the kinetic energy, the kinetic energy has been amplified by a factor of 72.9.  The Alfv\'{e}n Mach number, $M_A = \langle v/v_A\rangle$ can be simply determined as the square root of the ratio of the kinetic to magnetic energy.  At the peak of the kinetic energy, the turbulence is highly super-Alfv\'{e}nic, with $M_A \approx 230.$.  This property appears to be generally true as long as magnetic tension is not dominant.  

\begin{figure}[htb!] 
\epsscale{0.50}
\centering
\includegraphics[clip=true, scale=0.45]{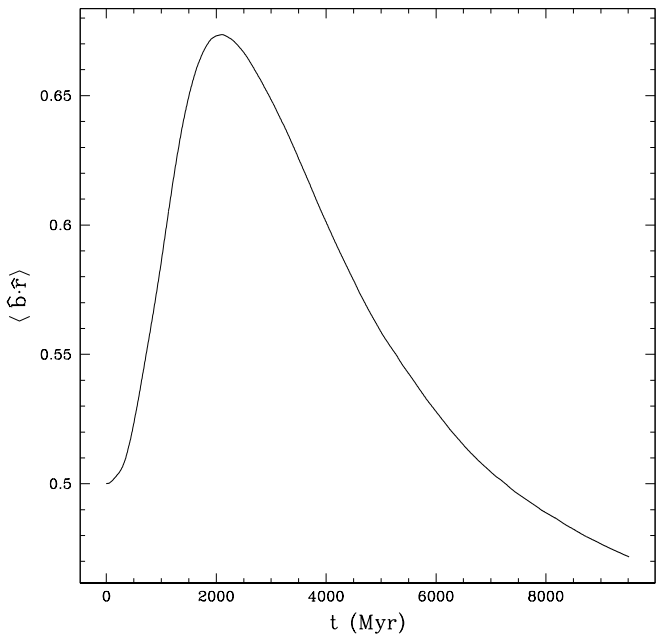}
\caption{The volume-averaged radial component of the magnetic field, \bdotr, is amplified even for the tangled magnetic field in run A3.}\label{fig:A3-bdotr}
\end{figure}

It is quite intriguing that the re-orientation of the magnetic field geometry can be observed quite vividly even for the tangled field case.  Starting with an initial geometry with \bdotr $=0.5$, the radial component is amplified to a peak value of 0.67 at 2.1 Gyr, before decaying after the free energy is exhausted.  Figure \ref{fig:A3-bdotr} shows the evolution of this quantity over the simulation.  
Simultaneously, we examine the evolution of the heat fluxes in this problem in the same way we did for run A2.  Clearly, the initial conductive heat flux is already a modest fraction of Spitzer due to the isotropically tangled magnetic field lines.  As the simulation evolves, the conductive heat flux normalized to the instaneous fiducial heat flux reaches a value of $f_{\textrm{Spitz}}=0.45$ at the peak of \bdotr.  The radial component continues to rise further past 0.5 as the temperature gradient is further depleted in this simulation.
\begin{figure}[htb] 
\epsscale{0.50}
\centering
\includegraphics[clip=true, scale=0.45]{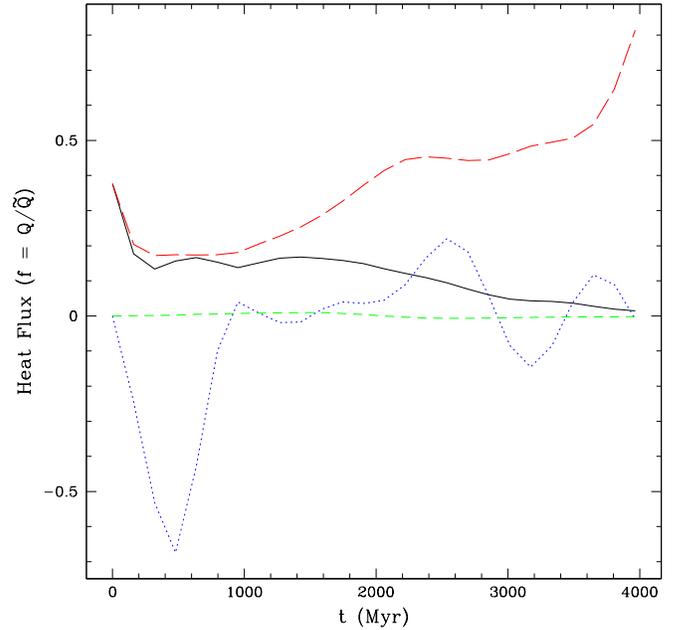}
\caption{This figure demonstrates the evolution of the radial heat fluxes for run A3 in a 100 kpc shell located 450 kpc from the cluster center.  The conductive heat flux normalized to the \textit{initial} fiducial heat flux (\textit{black, solid line}) reaches a maximum before decreasing, while normalized to the \textit{instantaneous} fiducial heat flux (\textit{red, long-dashed line}) continues to increase.  The heat fluxes due to mass advection (\textit{blue, dotted line}) and convection (\textit{green, short-dashed line}) are shown normalized to the initial fiducial heat flux.} \label{fig:A3-heatflux}
\end{figure}
Again the convective heat fluxes are quite small in both an instantaneous and time-average sense.  These simulations in a tangled field geometry prove the robustness of the MTI in tangled fields when the power is initially on a variety of scales. 
\section{Observational Consequences and Discussion}\label{sec:discuss}
We have shown in the previous section that properly accounting for anisotropic thermal conduction in the intracluster medium drives unique physics with potential observational consequences.  The resultant physics is significantly different from the pure MHD case without conduction or the isotropic conduction case.  Of course, without heating or cooling mechanisms included, the simulations presented here are insufficient to make predictions for any particular cluster.  Instead, we briefly outline here two potential observational consequences that will be a focus of future work.
\subsection{Turbulent Velocities}\label{subsec:discuss:velocities}
In this work, we predict rms turbulent velocities of order 3--4\% the speed of sound and peak velocities of 30--40\% of the sound speed.  In a relaxed cluster that has not undergone a major merger, this may be the dominant velocity component in the ICM; although, certainly wakes from passing galaxies may also be a significant contribution.  The typical ICM velocities for the MTI are predicted to be approximately an order of magnitude larger than the velocities that would result from pure isotropic conduction alone.  

To determine the measurability of these hydrodynamic motions, we must compare the Doppler broadening due to hydrodynamic motions to the simple thermal Doppler broadening.  The most success will come in measuring hydrogen- and helium-like iron lines since the thermal broadening is much smaller for the heavy iron nuclei.  We can express the ratio of the broadening mechanisms as
\begin{equation}
\frac{\Delta \nu_{\textrm{turbulent}}}{\Delta \nu_{\textrm{thermal}}} =
0.46 \left(\frac{v_{rms}}{60 \,\textrm{km}\textrm{s}^{-1}}\right)
\left(\frac{k_B T}{5 \,\textrm{keV}}\right)^{-1/2}
\left(\frac{m_i}{56 \,m_H}\right)^{1/2}.
\label{eqn:broadening}
\end{equation}
These two contributions are roughly equal for an rms turbulent velocity of 60 km s$^{-1}$ and a temperature of 1 keV.  For the peak velocity fluctuations, the hydrodynamic motions dominate; however, at the rms level, the velocity fluctuations will be harder to see.  Clever modeling of the actual line shapes, as in \citet{sun03}, may improve the detectability threshold of hydrodynamic turbulence.  

As of yet, there have been no successful measurements of ICM velocities.  Had the X-ray calorimeter on \textit{Astro-E2} (\textit{Suzaku}) not failed, it would have been able to image the velocity fluctuations of the ICM at the level of the thermal or hydrodynamic broadening.  The next chance for confirming any of these predictions will occur with \textit{Constellation-X}.  Its spectral resolving power is energy-dependent, but based on current performance requirements, should be able to resolve the predicted level of velocity fluctuations in iron lines with energies around 6 keV.  
\subsection{Magnetic Field Geometry}\label{subsec:discuss:fields}
Several authors have noted a discrepancy in the measurements of cluster magnetic fields, finding the fields derived from rotation measure (RM) observations are typically a factor of four to ten higher than fields derived from inverse Compton (IC) measurements \citep{pet01, ct02}.  The efforts at explaining this discrepancy are not particularly satisfactory.

First, consider fields derived from a rotation measure.  The RM can be written as
\begin{equation}
\mathrm{RM} = 812 \int_{0}^{L} n_e \boldsymbol{B}\cdot\boldsymbol{\dif l}\;\mbox{radians}\,\mbox{m}^{-2},
\label{eqn:clust:RM}
\end{equation}
where $n_e$ is given in cm$^{-3}$, $\boldsymbol{B}$ is measured in $\mu$G, and $\boldsymbol{\dif l}$ is measured in kpc \citep{ct02}.  The rotation measure is proportional to the component of the magnetic field along the line of sight times the electron density.  Typically, the field geometry is assumed to be isotropic in direction.  An enhancement in the radial field along the line of sight would represent an enhancement in rotation measure, and an overestimate of the magnetic field strength.

Alternatively, the magnetic field may be derived from inverse Compton emission from the CMB off relativistic electrons.  While the inverse Compton emission has no direct dependence on the magnetic field, the magnetic field strength may nonetheless be derived by comparing the inverse Compton emission to the synchrotron emission, provided they are both produced by the same population of relativistic electrons.  More illumination is provided by examining the formulae for the luminosity of both phenomena:
\begin{eqnarray}
L_{sync}& = &\frac{4}{3}\beta^2\gamma^2 c \sigma_T U_B, \\
L_{IC}& = &\frac{4}{3}\beta^2\gamma^2 c \sigma_T U_{\gamma},
\label{eqn:clust:IC-lum}
\end{eqnarray}
where $\beta=v/c$, $\gamma=(1-\beta^2)^{-1/2}$, $\sigma_T$ is the Thompson scattering cross section, $U_B=B^2/8\pi$ is the magnetic energy density, and $U_{\gamma}$ is the energy density of the photon field.  The similarity of these two equations is quite evident, giving
\begin{equation}
\frac{L_{IC}}{L_{sync}} \propto \frac{U_{\gamma}}{U_{B}}.
\label{eqn:clust:IC-rat}
\end{equation}
The inverse Compton emission peaks in the X-ray around 20 keV, corresponding to electrons with a $\gamma\sim 5000$.  The corresponding synchrotron emission peaks in the radio around 100 MHz \citep{bag98}.  

The IC emission is entirely independent of the magnetic field geometry; however, the synchrotron emission is a weak function of magnetic field geometry. Clearly, more detailed analysis will be required in the future; however, the basic idea is intriguing.  By simply positing a magnetic field that is preferentially radial to the line of sight, the rotation measure--derived field is overestimated while the IC/synchrotron--derived magnetic field is not significantly changed.  Future work will examine whether the action of the MTI can explain this observational discrepancy.
\subsection{Temperature Profiles} \label{subsec:temp}
It may seem as if the temperature profile is the most obvious observerable, however we have a ``chicken and egg" problem.  Our initial conditions posited in \S\ref{subsec:method:IC} reflect our observations of clusters \textit{today}.  Thus what we are observing are relaxed clusters that could be in the saturated state of any instability that has taken place.  In fact, the initial conditions for a relaxed cluster today should probably be taken from a typical disturbed cluster (e.g. Perseus) that has recently undergone a major merger.  Additionally, we have at this point neglected heating sources due to minor mergers, AGN activity, and infall that can be important in heating a cluster.  Nonetheless, we find that significant evolution of the temperature profile can occur in the relevant time between cluster major merger events.  In future work, we will work to more systematically connect our predicted temperature profiles with observations.
\subsection{Thermal Evaporation and Missing Baryons} \label{subsec:discuss:missing-baryons}
There have been several recent suggestions that Sunyaev-Zel'dovich (SZ) and X-ray measurements of galaxy clusters are incompatible.  As mentioned before, the X-ray measurements are sensitive to the line integral of the square of the density; whereas, the SZ signal is directly related to the line integral of the pressure (or thermal energy) of the baryons.  Recently, \citet{afshordi07} have used composite WMAP data to find that $35 \pm 8$\% of the cluster baryons are missing from the ICM. Namely, the authors argue that the gas fraction of clusters is consistently poor relative to the universe as a whole.  The validity of this measurement will likely be confirmed or refuted in the near future as much more accurate SZ surveys, including the Atacama Cosmology Telescope (ACT) and the South Pole Telescope (SPT) as well as numerous others, begin to make maps and measurements of clusters.  

This piece of observational evidence has motivated proposals for thermal evaporation of galaxy clusters, such as the work by \citet{loeb07}.  This work supposes that the outer portions of galaxy clusters could thermally evaporate, losing up to 10\% of their mass.  The primary proposed mechanism for this is leakage of suprathermal particles with large Coulomb mean free paths.  While this work is perhaps not theoretically robust, it does require magnetic fields that are preferentially radial in direction in order for this large mean free path argument to translate into suprathermal particles escaping.  Radial magnetic fields are of course one of the predictions of the nonlinear MTI simulations to date, thus satisfying one key requirement for this process.  Reconsidering this problem with a methodology similar to Parker's treatment of the solar wind could place this process on a more firm theoretical footing.  Of course, observations and theory are not yet well-developed on this topic, but it is worth keeping these facts in mind until the observations become more certain.
\section{Summary and Future Work}
The intracluster medium is a dilute magnetized plasma in which the mean free path is a moderate fraction of the size of a cluster.  In this plasma, heat flows exclusively along magnetic field lines making the ICM susceptible to the magnetothermal instability in regions where the temperature profile is decreasing outwards.  We have performed fully three-dimensional time-dependent MHD simulations of the intracluster medium and outlined its evolution and potential observational consequences.  We have also performed simulations without conduction and with only isotropic conduction to provide a basis for comparison and understanding of the MTI.  

There are several common features of the evolution of the MTI in the intracluster medium.  First, the MTI is capable of rearranging the temperature profile of the ICM on timescales significantly shorter than the Hubble time.  Second, the MTI drives subsonic convection that amplifies the magnetic field in a dynamo.  Third, the magnetic field geometry is driven to a highly-radially biased field.  Finally, we have self-consistently calculated $f_{\textrm{Spitzer}} \gtrsim 0.4-0.5$ for the first time and find that conduction can be very efficient at transporting energy.  Meanwhile, convective heat transport by the MTI in clusters is not efficient at all.  These conclusions simultaneously help understand several puzzles surrounding the ICM including the general flatness of the large-scale temperature profiles and the origin of the present-day cluster magnetic fields.  

We have also discussed several potential observational consequences of the MTI and anisotropic transport.  First, future X-ray satellites such as \textit{Constellation-X} may be able to image the subsonic velocity fluctuations.  Second, future work may explain the discrepancy between inverse Compton and rotation measure determinations of the magnetic field strength.  Finally, a better understanding of the masses of galaxy clusters may be obtained by understanding the ICM physics in much more detail.  Careful measurements of galaxy cluster masses can provide detailed cosmological tests.  

An important lesson from this work is that high--$\beta$ magnetic fields, though weak in terms of pressure, are not at all negligible.  In the long mean free path regime, such as in the intracluster medium, the physics of anisotropic transport along magnetic field lines can self-consistently drive the temperature profile evolution, magnetic field amplification and \emph{magnetic field geometry}, as well as turbulence.  Treating the magnetic field as a simple passive contaminant in hydrodynamic turbulence misses key relevant physics.

The future work here is manifold.  In the very near future, we will shift our attention inward to the cluster cores.  This region of the cluster, in which the temperature increases outward, is unstable to a cousin of the MTI known as the heat-flux-driven buoyancy instability, or HBI \citep{quat08, pq08}.  In contrast to the MTI, this instability drives field lines to be perpendicular to the background temperature gradient, potentially exacerbating the cooling flow problem.  In the short term, we will study the effects of the HBI and cooling in cluster cores.  For the global MTI calculations in particular, several steps remain.  First, a study of the cores of clusters including heating and cooling processes will likely require static mesh refinement.  Second, we have yet to do a thorough parameter study.  In such a study, we would quantify the variation of the MTI physics as parameters such as the cluster mass and central temperature are varied.  This type of understanding is very important for the use of clusters as standard candles in cosmology.  Finally, we are very interested in exploring the observational consequences of the MTI in more detail.  In particular, we would like to understand theoretically the modification of the magnetic field measurement through inverse Compton and rotation measure methods, as the geometry of the magnetic field changes from isotropic to radially-biased.  With a theoretical understanding, we could compute exact IC and rotation measure maps from a high-resolution simulated cluster for comparison with observational data.  

In short, clusters of galaxies provide a rich physics laboratory for testing Braginskii-MHD, the MTI, and many other phenomena.  We have developed analytical and computational tools for understanding these phenomena, and we will continue to apply these extended MHD models to improving our understanding of clusters of galaxies.  
\acknowledgements
We thank Eliot Quataert, Prateek Sharma, and Ben Chandran for many useful conversations.  We also acknowledge useful suggestions from Chris Reynolds, Eve Ostriker, and especially Jack Hughes.  I.~J.~P. is supported by NASA through the Chandra Postdoctoral Fellowship grant PF7-80049 awarded by the \textit{Chandra} X-Ray Center, which is operated by the Smithsonian Astrophysical Observatory for NASA under contract NAS8-03060.  I.~J.~P. was previously supported by the Department of Energy Computational Science Graduate Fellowship.  J.~M.~S. acknowledges support by the DOE through grant DE-FG52-06NA26217. This research was supported in part by the National Science Foundation through TeraGrid resources provided by the National Center for Atmospheric Research and the Pittsburgh Supercomputing Center.  Parts of this work were also performed on the Orangena and Della supercomputers at Princeton University.
\bibliography{ms}
\end{document}